\documentclass[runningheads]{llncs}

\usepackage{amssymb,amsmath,epsf}
\DeclareMathOperator{\dom}{\mathrm{dom}}
\DeclareMathOperator{\com}{\mathrm{com}}
\newcommand{\gisela}{Gisela}
\newcommand{\te}[1]{\text{{#1}}}

\begin{document}
\setcounter{page}{111}
\title{Enhancing Usefulness of Declarative Programming Frameworks through Complete Integration}
\titlerunning{Enhancing Usefulness of Declarative Programming Frameworks\ldots{}}
\author{G{\"o}ran Falkman\inst{1} \and Olof Torgersson\inst{2}}
\authorrunning{G. Falkman, O. Torgersson}
\institute{Department of Computer Science, University of Sk{\"o}vde,\\
           PO Box 408, SE--541 28 Sk{\"o}vde, Sweden\\
           \email{goran.falkman@ida.his.se}
\and       Department of Computing
           Science, Chalmers University of
           Technology and G{\"o}teborg University, SE--412 96 G{\"o}teborg, Sweden\\
           \email{oloft@cs.chalmers.se}}

\maketitle

\addtocounter{footnote}{1}
\footnotetext{In Alexandre Tessier (Ed), proceedings of the 12th International Workshop on Logic Programming Environments (WLPE 2002), July 2002, Copenhagen, Denmark.\\Proceedings of WLPE 2002: \texttt{http://xxx.lanl.gov/html/cs/0207052} (CoRR)}

\begin{abstract}
The Gisela framework for declarative programming was developed
with the specific aim of providing a tool that would be useful for
knowledge representation and reasoning within real-world applications.
To achieve this, a complete integration into
an object-oriented application development environment was used.
The framework and methodology developed  provide two alternative
application programming interfaces ({\sc api}s): Programming using
objects or programming using a traditional equational declarative
style. In addition to providing complete integration, Gisela also
allows extensions and modifications due to the general computation
model and well-defined {\sc api}s.  We give a brief overview of
the declarative model underlying Gisela and we present the
methodology proposed for building applications together with some
real examples.
\end{abstract}



\section{Introduction}
\label{s:introduction}
Today, the difference in availability and
quality of tools and libraries aimed at declarative programming
languages compared to what exist for, e.g., Java$^{\te{TM}}$ is
striking, to say the least. The non-declarative languages are often
\emph{good enough}, and the presence of mature and extensive
libraries and a variety of development tools simply outweigh the
advantages of using a declarative approach. Until the declarative
languages manage to close the gap, motivating the use of a
declarative language for general purpose, large scale programming
projects is hard, even though the language might have many
desirable properties for the task at hand.

An alternative approach then, is to make use of all the
development years put into legacy programming tools, and to combine
these with declarative programming. Thus, each programming
paradigm can be used for the task it does best. In this manner,
knowledge representation, reasoning, and other inherently
declarative activities can be programmed in a natural high-level
declarative way, and graphical user interfaces ({\sc gui}s),
network communication, or database access using other techniques.
A problem with this integration of different paradigms and tools
is that connecting the different parts of a system often is rather
complicated, again lessening the chance that declarative languages
really become used in real-world interactive applications.

In the Gisela project, we have taken this integrative approach to
making declarative programming more useful in the development of
real-world applications with {\sc gui}s. However, instead of providing some kind
of foreign-language interface, we have developed a system and
methodology for \emph{complete integration} of the different
programming paradigms used. Accordingly, Gisela is
not intended to be yet another declarative programming language,
but rather a general \emph{framework} for building embedded
declarative reasoning components within applications. As such Gisela provides:
\begin{enumerate}
\item A declarative computation model based on an abstract notion of a definition
\item An object-oriented framework
providing two different {\sc api}s: programming using objects and
`traditional' equational declarative programming
\item The
possibility to \emph{experiment} with definitional programming,
due to a general description of computations and the possibility
to introduce new classes into the object-oriented framework or
subclass existing ones.
\end{enumerate}
The result is a seamless integration with an object-oriented
programming environment and classes for
development of desktop and web applications, which is
open for modifications and extensions through such concepts as
inheritance and subclassing. From the point of view of systems
development, this is just as important as seeking the perfect
computation model or fastest implementation.

Gisela evolved out of the need for a declarative programming tool
suitable for the development of real-world applications, and it has been used in the development of a
knowledge-based system in the area of oral medicine
\cite{alifalhaljonnaztor:mie00,padl02}.

The rest of this paper is organized as follows. In Sect.
\ref{s:definitional_programming}, the declarative model of Gisela
is outlined and an example using traditional declarative
programming is given. Section \ref{s:complete_integration}
explains how declarative and object-oriented programming are used
together. Section \ref{s:example_applications} gives a few example
applications. The paper is concluded in Sect. \ref{s:discussion}
with a general discussion.



\section{Definitional Programming}
\label{s:definitional_programming}

A common concept of declarative programming is the concept of a \emph{definition}.
Function definitions
are given, predicates are defined etc. However, focus is on what
we define, on the functions and predicates respectively. \emph{Definitional programming}
is an approach to declarative
programming where the definition is the basic notion, i.e., focus
is on the definitions themselves, not on what we define.

In this section we outline the model for computing with
definitions, which forms the basis of \gisela. A complete presentation is given in \cite{tor:thesis}.

\subsection{Definitions}
\label{ss:definitions}

The definitional model of Gisela can be seen as an extension
of logic programming based on the theory of \emph{partial
inductive definitions} \cite{pid}.
In this theory, a definition $D$ is given by
\begin{enumerate}
\item two sets: the \emph{domain} of $D$, written $\dom(D)$, and the \emph{co-domain} of $D$,
written $\com(D)$, where $\dom(D) \subseteq \com(D)$,
\item and a \emph{definiens} operation: $D : \com(D) \rightarrow {\cal P}(\com(D))$.
\end{enumerate}
Objects in $\dom(D)$ are called \emph{atoms} and objects in
$\com(D)$ are called \emph{conditions}.

A natural presentation of a definition $D$ is that of a system of equations
$$
\begin{array}{lcl}
D & \left\{ \begin{array}{lcl}
        a_{0} & = & A_{0} \\
                  & \vdots & \\
        a_{n} & = & A_{n} \\
                  & \vdots &
        \end{array}
    \right.
& \quad n \geq 0\enspace,
\end{array}
$$
where $a_{0},\ldots \in \dom(D)$ and $A_{0},\ldots \in \com(D)$, i.e., all pairs
$(a_{i},A_{i})$ such that $a_{i} \in
\dom(D)$ and $A_{i} \in D(a_{i})$. Note that an equation $a = A$ is just a notation for $A$ being a
member of $D(a)$.

\subsection{Programs and Computations}
\label{ss:programs_and_computations}

Programs consist of \emph{data definitions}, \emph{method definitions}, and \emph{state definitions}.

Data definitions describe the declarative content of a program. As an example, pure Prolog is a
subset of definitional programming \cite{halsch:ptaI}, and using the simple interactive system of
Gisela (see Sect. \ref{ss:simple_ide}),
a data definition \verb+permutation+ with two Prolog-style predicates can be given:
\begin{verbatim}
definition permutation.

perm([],[]).
perm([X|Xs],[Y|Ys]) = select(Y,[X|Xs],Zs), perm(Zs,Ys).

select(X,[X|Xs],Xs).
select(Y,[X|Xs],[X|Ys]) = select(Y,Xs,Ys).
\end{verbatim}

A computation is a transformation of an initial state definition
into a final state definition (referred to as a \emph{result
definition}). In addition to the result definition, the result of
a computation contains an answer substitution for variables in the
initial state definition. Result definitions contain information
about how they were computed. If we are not interested in the
complete structure of a result definition, it can be simplified in
different ways.

In this example, the result definition is of no particular interest so
the system is set to display answer substitutions only:
\begin{verbatim}
G3> restype(vars_only).
\end{verbatim}

Method definitions describe algorithms, or search
strategies, used to compute solutions. The method definitions give the computation steps that transform
the initial state definition into the result definition.

Continuing with the example, we define a method definition \verb+prolog+:
\begin{verbatim}
method prolog(P).

prolog = [] # some r:matches(true).
prolog = [prolog, r:P] # all not(r:matches(true)).
\end{verbatim}
Parameterized with the data definition \verb+P+, \verb+prolog+
computes on the right-hand side of the state
definition as long as the right-hand side does not equal \verb+true+.
When the right-hand side equals \verb+true+ the computation stops.

A \emph{query} is the application of a method definition to an
initial state definition. To get a Prolog query, the initial state
definition is set up with the logic programming goal as the single
right-hand side. In this example, we instantiate \verb+prolog+
with \verb+permutation+ and
 apply the result to \verb+{true = perm([a,b,c],L)}+:
\begin{verbatim}
G3> prolog(permutation){true = perm([a,b,c],L)}.
L = [a,b,c] ? ;
L = [a,c,b] ?
yes
\end{verbatim}

The operational semantics is given by a calculus \cite{tor:thesis}, in which the inference
rules interpret the conditions used in method definitions. The computation model has to handle
certain choices, e.g., the order in which equations should be considered, the ordering of the
elements of the definiens, and how to transform result definitions. These choices are handled
by an \emph{observer}. The default observer considers equations in the order they are defined
in the program and allows three types of transformations of the result definitions. In this way, the notion
of an observer provides a `hook' into the computation model.

Logic programming is but one way to use definitional programming. Other techniques
are studying properties of definitions, such as similarity
\cite{falkman:ewcbr2k} and separated definitional programming
\cite{fal:progsep}.



\section{Complete Integration}
\label{s:complete_integration}

Gisela was developed with the specific aim of
providing a tool that would be useful for knowledge representation
and reasoning within real-world, state-of-the art, desktop and
web applications. Due to the presence of very good object-oriented
tools for application development it was decided that the most
practical approach would be to provide for a seamless integration
between these tools and Gisela instead of trying to build a
definitional {\sc gui} programming library.

Gisela is currently integrated with the OpenStep framework \cite{openstepspec},
which means that Gisela is implemented in Objective-C, an
object-oriented extension to {\sc ansi} C. For the future, we are considering porting Gisela to
a more widespread platform, e.g., Java$^{\te{TM}}$, for the obvious reason of greater platform
independence.

Gisela adheres to the common use of \emph{libraries} (\emph{frameworks} in OpenStep,
\emph{packages} in Java$^{\te{TM}}$) for providing and encapsulating
all the functionality for a certain
task. A framework should be complete enough to be used as is, but
also flexible enough to allow modification through subclassing.
Building an application in a framework-based environment is just
to provide the application specific details, all the general
machinery and design patterns are set up by the frameworks used.

From this perspective, the aim of Gisela is to be just another
framework, a framework providing all the necessary tools for embedding
definitional programming components into applications.
The situation can be compared to the use of a
relational database in an application: In such an application,
a database framework is used to seamlessly connect the application to an
existing database engine using classes in the framework. All the
details of, e.g., {\sc sql}, is handled by the framework, and if needed
the framework can be extended through subclassing. In an
application using Gisela, a declarative subsystem can be
integrated through the use of classes in the framework, extending
them if necessary. Furthermore, a traditional equational {\sc api}
can be used to program the declarative part if desired, and the
framework provides everything that is necessary to load and
execute the definitional part at runtime.

\subsection{The Gisela Framework}
\label{ss:gisela_framework}

In principle, Gisela represents each entity involved in a
computation by an object of a corresponding class. Thus, an
equation is represented by an object of the \verb+DFEquation+
class and so on. Additionally, the framework defines a number of
interfaces (\emph{protocols} in Objective-C), which declare the
functionality of each kind of object. Thus, apart from
subclassing, it is possible to replace classes by new ones
implementing the required interfaces.

Internally, Gisela consists of three frameworks: \verb+DFDefinitions+,
in which data definitions are implemented,
\verb+DFMethods+, which implements all classes needed to construct method definitions,
and \verb+DFComputing+, which implements the classes for performing computations.
The separation enables definition classes to used by their own and keeps the frameworks reasonably
sized.

\subsubsection{Computing Machinery.}
\label{sss:computing_machinery}

The heart of the Gisela framework is the \emph{D-Machine},
implemented by the \verb+DFDMachine+ class. The D-Machine performs the
actual computations by implementing the calculus in the underlying definitional model.
The D-Machine only relies on a few well-defined operations on data and
method definitions, opening up for different ways to realize these
parts.

The most important methods of the D-Machine interface are:
\emph{init}(\emph{Delegate},\ \emph{Observer}), \emph{setQuery}(\emph{Query}),
\emph{nextAnswer}, and \emph{allAnswers}.

A D-Machine may be set up in different ways depending on the
context. The machine
can run in the same thread as the object creating it or in a
separate thread, which might be more appropriate for interactive
applications. The machine's observer can be any object
implementing the appropriate methods. The delegate is an object
which handles certain things for the machine and receives
notifications at times. It can be the same as the observer or
another object.

The observer interface declares methods which provide hooks into
the D-Machine (see Sect. \ref{ss:programs_and_computations}), e.g.,
\emph{transformResult}(\emph{ResultDefinition},\ \emph{ResultType}).

\subsubsection{Data Definitions.}
\label{sss:data_definitions2}

All data definition classes must implement the methods
in the \verb+DFDefinition+ interface, of which the most important are:
\emph{inDom}(\emph{Object}) and \emph{inCom}(\emph{Object}),
which check if \emph{Object} is a member of
the domain or co-domain respectively, and \emph{def}(\emph{Object}),
which computes the definiens of \emph{Object}.

For a data definition class to be valid, the definitions of
the above methods should implement the behavior given by the abstract
description of a data definition given in Sect. \ref{ss:definitions}.

The D-Machine treats data
definitions as black boxes. All it knows about data definitions is
that a data definition can be used to find the definiens of an object
and that there may, in general, be more than one result.

In the general case, computing the definiens of an object with
respect to a data definition is a complex operation. Therefore,
the framework provides several
specialized data definition classes handling various simpler
definitions. The most common classes are \verb+DFModifiableDefinition+,
which implements common behavior of mutable definitions, and
\verb+DFMatchingDefinition+, which is suitable when matching, and not unification,
should be applied in the definiens operation.

\subsubsection{Method Definitions.}
\label{sss:method_definitions2}

What is special about method definitions is that
the definiens operation is always performed with respect to a
given state definition. Therefore, the class \verb+DFMethod+ (a subclass of
\verb+DFModifiableDefinition+) also declares a method
\emph{defWithStateDefinition}(\emph{StateDefinition}).

As with other entities involved in computations, users using the \gisela\
frameworks may implement new method definition classes as long as
they implement the \verb+DFMethod+ interface. Of course, subclassing
is also possible.

\subsection{Building an Application}
\label{sss:building_an_application}

From Gisela's point of view, an application consists of the application binary
and a resources folder, the latter containing all sorts of resources needed by
the application. The items in the resources folder can dynamically be loaded into
the application at run time. This
means that it is straightforward to have text files in the
resources folder representing definitions and computation methods.
These text files can be parsed into definition objects at run time
using the {\sc api} provided by the \gisela\ framework.
Consequently, any \gisela\ program developed using equational
presentations can smoothly be integrated into an application.

\subsubsection{Model--View--Controller.}
\label{sss:model_view_controller}

The general design approach used when integrating Gisela into an
application is the Model--View--Controller ({\sc mvc}) paradigm.
{\sc mvc} is a commonly used object-oriented software development
methodology. When {\sc mvc} is used, the \emph{model}, i.e., data
and operations on data, is kept strictly separated from the
\emph{view} displayed to the user. The \emph{controller} connects
the two, and decides how to handle user actions and how data
obtained from the model should be presented in the view. Applied
to the \gisela\ setting, the model of what an application should do is
implemented using definitional programming in \gisela. The view
displayed to the user can be of different kinds, desktop
applications, web applications etc. In between the view presented
to the user and the \gisela\ machinery there is a controller
object, which manages communication between the two parts. One
advantage of this approach is, of course, that different views may
be used without changing the model.

The general methodology to use \gisela\ to build applications thus
becomes:
\begin{itemize}
\item Decide what definitions are needed for the definitional part of the application.
\item Write the syntactic representations of the \gisela\ program part and add the resulting files to the application's resources.
\item At run time, load the definitional resources into objects representing them and create the desired number of \verb+DFDMachine+ objects for running queries.
A \verb+DFDMachine+ object can be set up to interact with the
application in various ways, depending on the demands of the
application.
\item Build a \verb+DFQuery+ object, representing the query, from user input somehow.
\item Ask a \verb+DFDMachine+ to run the query represented by the \verb+DFQuery+ object.
\item Present the result, represented by a \verb+DFAnswer+ object, to the user somehow.
\end{itemize}

The \gisela\ approach is in line with the Model-View-Controller
paradigm, where \gisela\ is used to build the model, and the
controller and view are constructed using standard programming
tools and available libraries. Typically, when a new model object is created it
allocates a \verb+DFDMachine+ object and loads the required
definitional computation resources contained within the
application.

\subsection{Extending Gisela's Capabilities}
\label{ss:extending_gisela}

We show a couple of examples of how the Gisela framework can be modified
or extended by developers using it to build applications.

\subsubsection{A Simple Database Adaptor.}
\label{sss:simple_database_adaptor}

In the MedView project \cite{alifalhaljonnaztor:mie00}, results from a large number of
examinations are stored as text files in a knowledge base. The
conceptual view of the knowledge base is that of a collection of
definitions. However, the format of the files is not one that can
be directly parsed by any standard definition class. Thus, to be able to
manipulate MedView data, a first step is to extend the framework
with a new definition class, called \verb+MVTreeDefinition+,
which understands the used format.

From the D-Machine's point of view, objects of the
\verb+MVTreeDefinition+ class are no different from, e.g.,
definitions created from equational presentations like
\verb+permutation+ shown in Sect.
\ref{s:definitional_programming}.

\subsubsection{Another Observer.}
\label{sss:another_observer}

The observer is responsible for selecting the order in which
equations are considered. If we want to
consider the left-most equation only, we can
introduce a new observer class. In this class we override the
appropriate method from the default observer:
\begin{verbatim}
@implementation LeftMostObserver
- (NSArray *)selectEquationsWithWord:(DFWord *)aWord
                     stateDefinition:(DFStateDefinition *)stateDef
                            andHints:(NSArray *)hints {
     return [NSArray arrayWithObject:[NSNumber numberWithInt:0]];
}
@end
\end{verbatim}
The left-most equation is the one at index \verb+0+.

The power of
object-oriented programming in general, and inheritance in
particular, lets us experiment with definitional computations
using the \gisela\ framework as a basis.



\section{Example Applications}
\label{s:example_applications}

In this section, we illustrate the use of Gisela with a few
examples built using the integrative approach to application
development.

To further increase the usefulness of Gisela, some basic tools
supporting application development have been built. The development of the tools themselves
is based on the use of the Gisela framework in combination with
object-oriented programming to handle user interaction.

\subsection{A Simple {\sc ide}}
\label{ss:simple_ide}

Using the object-oriented {\sc api}s of Gisela and following the {\sc mvc} paradigm,
we have written a simple interactive system useful for developing
and testing definitional programs in a traditional way.

However, nowadays, one typically uses an {\sc ide} for management of code, application
resources, debugging etc., rather than just a text editor
and a shell. In the world of imperative and
object-oriented programming, there exists a large number of
high-quality {\sc ide}s. For declarative programming, {\sc
ide}s are scarce.

To simplify development of equational definitional programs, a basic {\sc
ide} is being developed. The general principle in the development
of the {\sc ide} is the same as in application development in
general: Object-oriented programming is used to build the {\sc
gui}, and this is then hooked up to Gisela to
manage definitions.

The {\sc ide} is centered around
the notion of a project. A project consists of a number of related
data and method definitions. The main window consists of a
toolbar with common commands at the top, an outline view showing
all the project's files to the left, and a main area to the right
for editing source code etc. The main area can be split into
several views for different tasks, e.g., running queries from
within the {\sc ide}. To parse a data or method definition,
classes provided by the framework are used. To run a query, a
\verb+GiselaRunTime+ object is used, however hooked up to a {\sc
gui} rather than a command line interface as above. The user can
easily control properties of the D-Machine using pop-ups.

Although rather incomplete, the {\sc ide} has been designed to be
extendable. For example, once a debugger has been developed it can easily be added.

\subsection{MedViewer -- A Real Application}
\label{ss:medviewer}

\begin{figure}[t]
\begin{center}
\setlength{\epsfxsize}{0.66666\textwidth}\leavevmode
epsfbox{medviewer.eps}
\end{center}
\caption{MedViewer data exploration tool}\label{fig:medviewer}
\end{figure}

MedViewer is an application for dynamic data exploration developed
within the MedView project. The user dynamically selects cases from a knowledge
base containing about 3000 patient examinations. The selected
cases are then visualized in different ways, e.g., using
scatter plots or bar charts (see Fig. \ref{fig:medviewer}).

MedViewer uses the Gisela framework in several ways. All data in
the knowledge base is represented by \verb+MVTreeDefinition+
objects. To make selections, definitional computations are used,
based on method definitions stored within the
application. The user can also construct groups of
related cases. These groups are data definitions developed by
users using a special {\sc gui}. Finally, to generate case
descriptions, data definitions are used to represent text
templates, using some custom data definition classes. The text
generator does not use a D-Machine, but hard-wired procedural
behavior, yet another way of using Gisela.



\section{Discussion}
\label{s:discussion}

Our experiences from using Gisela are positive: It provides
seamless integration with tools for building desktop and web
applications, and can easily be modified or extended when
functionality not present in the framework is needed. If desired,
the framework can also be used for declarative programming at
different levels. For instance, it is possible to use only data
definitions describing a domain and implement the procedural
behavior without using Gisela.

Gisela has been evaluated in practice within the MedView project.
Applications for knowledge acquisition, data presentation using
natural language generation, and information visualization have
been developed. These tools are in daily use at several clinics
attached to {\sc somnet} (Swedish Oral Medicine NETwork).

There are two things that set \gisela\ apart from most approaches
to declarative programming: (i) \gisela\ is not a general-purpose
programming language, rather it is a system for realizing a
certain set of definitional models and (ii) \gisela\ is a
framework with an abstract definition, specifically aimed at
allowing experiments and modifications within the general
definitional and computational models. Furthermore Gisela is
designed for complete integration into a surrounding
object-oriented environment.

There are  similar realizations of declarative languages,
typically targeting Java$^{\te{TM}}$. For instance, Frapp\'{e}
\cite{frappe} that implements the abstract concept of Functional
Reactive Programming \cite{haskell:hse} in Java$^{\te{TM}}$ in a
way related to how Gisela applications can be developed using the
object-oriented {\sc api}.   Another is {\tt tu}Prolog
\cite{tuProlog}, which implements Prolog in Java$^{\te{TM}}$ in a
manner enabling a tight integration between the two languages.
However, contrary to Gisela the declarative part is a fixed
programming language instead of an extendable framework.

An alternative to mixing declarative programming with other tools
 to build {\sc gui}s is to provide declarative libraries, e.g.,
\cite{curry:gui}. However, we have not seen any examples that are
on par with their object-oriented counterparts in terms of ease-of
programming, development tools, or quality of the resulting
interface. When it comes to interfacing declarative system with
other tools there are several choices. Most declarative languages
have foreign language interfaces, but these are often low-level
and rather cumbersome to use. A more high-level generic interface
is shown in \cite{padl02:eclipse}. Another way to achieve
integration, perhaps more related to Gisela, is to compile into a
common platform providing integration facilities. An example of
this is the use of the {\sc .net} common language runtime as
target for Mercury \cite{mercury:dotnet}. There are also several
systems that compile to Java$^{\te{TM}}$ bytecode, such as MLj
\cite{mlj}. At another level is providing and using interfaces for
standardized techniques for building component based systems such
as {\sc com/corba}, e.g., \cite{mcorba}.

The complete integration approach used in Gisela differs from all
of the above in that it does not require any kind of interface to
the system it is integrated into. There is a price to pay though:
The complete integration is with one particular object-oriented
application development environment, currently OpenStep - to
integrate with another one a new implementation is required.

Currently, we have no plans to extend Gisela to handle
sophisticated interaction with the user. Instead, we advocate the
integrated approach with a definitional model realized in \gisela\
and an interface part programmed using other, more suitable,
tools. Thus, the declarative framework, in effect, extends the
entire application development framework surrounding it. In our
experience, this is the most practical approach. Integration is
 one  one crucial factor for the success of declarative
programming. Another factor is making a shift from the development
of programming languages to systems development: In studying
object-oriented system development, learning a language like
Java$^{\te{TM}}$ is but a small part, learning modelling languages
and design patterns is equally important. \gisela\ is our attempt
at providing a useful declarative programming component.



\end{document}